\documentclass{new_tlp}
\usepackage{mathptmx}
\usepackage{graphicx}
\usepackage{xspace}
\usepackage{xcolor}
\usepackage{color}
\usepackage{amsmath}
\usepackage{listings}
\usepackage[hyperindex,breaklinks]{hyperref}
\usepackage{hhline}

\newcommand{\set}[1]{\mathbf{#1}}

\newcommand{\Name}{FormuLog\xspace}
\newcommand{\icode}[1]{\texttt{#1}\xspace}

  \title[\Name: Datalog for static analysis involving logical formulae]
        {\Name: Datalog for static analysis involving logical formulae}

  \author[A. Bembenek and S. Chong]
         {AARON BEMBENEK \\
         Harvard University, Cambridge, MA 02138, USA\\
         \email{bembenek@g.harvard.edu}
         \and
         STEPHEN CHONG \\
         Harvard University, Cambridge, MA 02138, USA\\
         \email{chong@seas.harvard.edu}}

\begin{document}

\maketitle

  \begin{abstract}
Datalog has become a popular language for writing static analyses.
Because Datalog is very limited, some implementations of Datalog for static analysis have extended it with new language features.
However, even with these features it is hard or impossible to express a large class of analyses because they use logical formulae to represent program state.
    \Name fills this gap by extending Datalog to represent, manipulate, and reason about logical formulae.
We have used \Name to implement declarative versions of symbolic execution and abstract model checking, analyses previously out of the scope of Datalog-based languages.
While this paper focuses on the design of \Name and one of the analyses we have implemented in it, it also touches on a prototype implementation of the language and identifies performance optimizations that we believe will be necessary to scale \Name to real-world static analysis problems. 
  \end{abstract}


\section{Introduction}

The logic programming language Datalog has become a popular domain-specific language for writing static analyses~\cite{reps,bddbddbUsing,doop,souffle2,flix}.
As a declarative language that embodies Kolwaski's principle of separating a program's ``logic'' (what it computes) from its ``control'' (how it computes it)~\cite{kowalski}, Datalog is a good fit for programming a static analysis, where often the logic of the analysis is substantially simpler than the control necessary to efficiently compute it.
In particular, many analyses logically consist of mutually recursive sub-analyses that in theory interact elegantly, but in practice can be tricky to coordinate.
Datalog makes it easy to state the dependencies found in these types of analyses, enabling analysis designers to focus on perfecting the logic of their analyses without worrying about the low-level control details.

Because Datalog is a very restricted language and there are many analyses that cannot easily be expressed in pure Datalog, implementations of Datalog for static analysis have extended the language with additional features. For instance, Souffl\'e adds a form of $n$-ary constructors~\cite{souffle2}, and Flix adds user-defined functions and support for reasoning over lattices~\cite{flix}.
However, there are still common analyses that cannot be naturally expressed even with these features: namely, analyses that reason about program behavior through logical formulae.
Such analyses include symbolic execution, abstract interpretation over the predicate domain, and various forms of model checking.
These analyses symbolically represent reachable program states as logical formulae, and then reason about them in theories that support concepts such as machine integers and arrays.
No Datalog variant currently provides language abstractions for computing over these types of formulae.

This paper presents a language called \Name that extends Datalog to support this class of analyses, and thus explores how the declarative benefits of Datalog can be extended to these analyses.
\Name makes it easy to represent, manipulate, and reason about logical formulae:
\begin{itemize}
    \item Logical formulae can be represented by ground (i.e., variable-less) terms formed from $n$-ary constructors, such as the term \icode{or(true, false)}, which encodes the formula $True \vee False$ using the binary constructor \icode{or} and the nullary constructors \icode{true} and \icode{false}.
    \item Logical formulae can be manipulated by user-defined functions that consume and produce ground terms.
        For example, a function might rename the logical variables in a formula.
    \item Logical formulae can be reasoned about through built-in functions that invoke an external satisfiability modulo theories (SMT) solver.
        For instance, the unary function \icode{is\_sat} evaluates to true if its argument is satisfiable when interpreted as a logical formula.
        A type system ensures that only terms that can be interpreted as formulae are applied to these special functions.
\end{itemize}
These features are combined in such a way that \Name remains a declarative language, in that the meaning of a \Name program is independent of how it is evaluated.
On the one hand, this means that analysis designers can reason about the correctness of their analyses without worrying about how the \Name runtime will actually execute them.
On the other hand,
this means that the \Name runtime is free to choose how to evaluate an analysis and can aggressively apply optimizations, such as rewriting the analysis and running it in parallel.

Languages like \Name can help address concrete challenges faced by modern static analyses.
For example, \citeN{beast} explain how state-of-the-art static analyses struggle to analyze programs that make heavy use of frameworks, such as Java web applications.
These type of applications are difficult to analyze because frameworks typically rely on dynamic language features like reflection.
Analyses cannot ignore these features because they play such a large role in framework behavior, but most analyses that treat these features precisely do not scale.
Because of this, Toman and Grossman suggest a hybrid approach: a scalable meet-over-all-paths analysis is used for the application code, and a variant of symbolic execution, being more precise but less scalable, is used for the framework code.
As control flow goes back and forth between the application and the framework, these analyses interact in a mutually recursive way.
This is the perfect setting for a Datalog-like language that makes it easy to encode interdependent analyses.
But, of course, that language must be able to effectively support the relevant analyses, including analyses such as symbolic execution that are currently not supported by Datalog variants.
\Name takes a step in this direction by making it easier to declaratively encode analyses that compute with logical formulae.

Section~\ref{sec:background} provides background on Datalog and how logic programming has previously been used for static analysis.
Section~\ref{sec:design} presents the design of \Name and Section~\ref{sec:cases} demonstrates how it enables a declarative encoding of symbolic execution.
Section~\ref{sec:prototype} discusses a current prototype of the language, gives some initial performance results, and sketches potential optimizations.

\section{Background and related work}\label{sec:background}

\begin{figure}[t!]
        \begin{align*}
            &\text{Variables} & X,Y,Z &\in \set{Var} \\
            &\text{Constructor symbol} & a, b &\in \set{CSym} \\
            &\text{Predicate symbol} & p &\in \set{PSym} \\
            &\text{Term} & s, t &::=~X \mid a \\
            &\text{Atom} & A &::=~p(t_1, \dots, t_n) \\
            &\text{Clause} & C & ::=~A~\icode{:-}~A_1, \dots, A_n.
        \end{align*}
    \caption{Datalog is a simple language that forms the basis of \Name.}
    \label{fig:datalog}
\end{figure}

Datalog is a simple logic programming language (Figure~\ref{fig:datalog})~\cite{neverdared,recursivequery}.
A Datalog program is defined by a set of clauses, where each clause consists of a single head atom and a set of body atoms.
An atom is a predicate symbol applied to a list of terms, where each term is either a variable or an uninterpreted constant (i.e., a nullary constructor).
Each predicate symbol is associated with either an extensional database (EDB) relation or an intensional database (IDB) relation.
An EDB relation is tabulated explicitly through facts (clauses with empty bodies), whereas an IDB relation is computed through rules, clauses that have non-empty bodies.

Datalog has an elegant logical interpretation that leads to a declarative semantics, in the sense that the meaning of a Datalog program does not depend on how it is evaluated. A clause
$$A~\icode{:-}~A_1, \dots, A_n.$$
can be interpreted as the logical formula
$$\forall \overline{X}.(A_1 \wedge \cdots \wedge A_n \rightarrow A)$$
where $\overline{X}$ is a vector of the variables that occur in the clause.
The clauses of the program can then be viewed as the axioms of a first-order theory, and the semantics of the program is the least Herbrand model of that theory
(Datalog puts syntactic restrictions on programs to guarantee that such a model exists and is finite).

\subsection{Datalog-based static analysis frameworks}

Subsequent to the early use of Datalog by~\citeN{reps} for deriving demand-driven versions of interprocedural analyses, a variety of static analysis frameworks have been developed that leverage Datalog.
A binary decision diagram-based implementation of Datalog, bddbddb~\cite{bddbddbUsing}, has been used to compute context-sensitive pointer analysis~\cite{bddbddbPointer} and taint analysis for Java applications~\cite{bddbddbSecurity}, and has also been used as one of the analysis languages for Chord, a Java bytecode analysis framework.%
\footnote{Chord is available at \url{https://bitbucket.org/psl-lab/jchord/}.}
The points-to analysis framework Doop uses a heavily-optimized version of Datalog to outperform contemporary state-of-the-art pointer analyses built on binary decision diagrams~\cite{doop}.
Instead of evaluating an analysis written in Datalog directly, Souffl\'e treats the analysis as a synthesis specification: from the Datalog program, it synthesizes a C++ implementation of that analysis~\cite{souffle2,souffle1}.
Souffl\'e extends Datalog with a type system and a form of $n$-ary constructors, and has been used to synthesize pointer and security analyses for Java applications.
Because the addition of $n$-ary constructors makes Datalog Turing-complete~\cite{recursivequery}, Souffl\'e is technically powerful enough to encode the analyses that \Name is designed for; however, it does not provide the abstractions to make this practical.

Flix extends Datalog with monotone functions written in a pure functional language, algebraic data types, and user-defined lattices~\cite{flix}.
Although \Name also has pure functions and algebraic data types, the two languages target different shortcomings in the use of Datalog for static analysis.
Flix addresses the fact that it is inefficient or impossible to state analyses in Datalog that compute over lattices that are not the powerset lattice.
\Name addresses the fact that Datalog provides no way to manipulate and reason about logical state.
That being said, \Name would benefit from some type of aggregation, and the lattice approach of Flix might be a good fit.
A recent paper demonstrates the implementation of dataflow analysis in Datafun, a language that combines Datalog and higher-order functional programming~\cite{datafun}.
By interweaving Datalog with functional programming, these languages are related to functional logic programming~\cite{flp}.

There has also been recent interest in automatically refining the abstractions used in Datalog analyses~\cite{datalogabsrefine} and in example-based synthesis of analysis implementations in Datalog~\cite{datalogsynthesis}.

\subsection{Constraint logic programming}

Constraint logic programming (CLP) extends logic programming with constraint solving over various theories~\cite{clp,clpsurvey,clp25}.
Because they both provide mechanisms for reasoning about logical formulae,
there are some similarities between \Name and CLP.
However, \Name takes a substantially different approach to constraint solving than traditional CLP, in which constraints are represented by distinguished predicates.
When the CLP runtime encounters one of these predicates during evaluation, it adds the relevant constraint to its constraint store and checks that the store is still consistent.
This means that a constraint is evaluated in the context of all the constraints that have come before it, and there is no way to manipulate, propagate, or reason about constraints outside of this mechanism. 
On the other hand, by reifying logical constraints as terms and only treating them as constraints when applied to certain functions, \Name provides the programmer with more control over how constraints should be handled.
This flexibility is necessary for some static analyses.
For instance, model checkers in the style of BLAST~\cite{lazyabs} dynamically associate logical formulae with individual program points (that is, code locations in the program under analysis).
These analyses must be able to explicitly refer to and manipulate the formula associated with a given program point, and might want to reason about a formula in a local (non-global) context.
It would be hard to meet these requirements using a traditional CLP approach in which constraints are represented by predicates and interpreted within a global store.

Some CLP-like systems do represent constraints as terms~\cite{lpsat,clpqr}, and some of them have been used to implement static analysis algorithms such as abstraction refinement model checking~\cite{armc}.
Interestingly, \Name represents constraints using {\em ground} terms, while these systems represent constraints using (in general) {\em unground} terms, since in these systems constraint-level variables are CLP-level variables.
In program analysis, where we want to reason about constraints over variables in the input program, this has the effect of punning logic programming-level variables with input program variables.
While this has some nice benefits, such as being able to use unification to avoid explicitly renaming variables in formulae~\cite{armc}, it does require additional bookkeeping by the analysis to ensure that constraints are formed over the correct variables.
As this bookkeeping involves propagating information about logic programming-level variables, it seems difficult to use this approach in a Datalog-like setting, where Datalog's range restriction does not allow a fact that has an unbound variable to be derived (and thus propagated). 

Furthermore, since many CLP implementations are extensions of Prolog~\cite{clp25}, they inherit the limitations of Prolog's depth-first search evaluation strategy.
For example, the termination of an analysis written in one these CLP languages is sensitive to the order of clauses and the order of atoms within a clause body, a problem not faced by Datalog.
Additionally, since they are not as tightly coupled to a particular evaluation strategy, Datalog-based systems like \Name are freer to apply optimizations that can help scale static analyses, such as analysis rewriting~\cite{doop} and parallelization~\cite{souffle2}. 

\section{Language design}\label{sec:design}

The goal of \Name is to make it possible to represent, manipulate, and reason about logical formulae.
To achieve this goal, \Name extends the grammar of Datalog terms with $n$-ary constructors and function calls, and imposes a type system to enforce the correct use of these more complex types of terms.
Section~\ref{sec:extensions} introduces \Name's approach to these features, which are also present, to one extent or another, in previous variants of Datalog for static analysis~\cite{souffle2,flix}.
Section~\ref{sec:formulae} describes how \Name combines these features with built-in support for SMT solving to enable computation involving logical formulae. 
Section~\ref{sec:semantics} concludes with a short discussion of the semantics of \Name.

\subsection{Types, constructors, functions, and relations}\label{sec:extensions}

\begin{figure}[!t]
    \begin{verbatim}
define type 'A tree = leaf | node('A tree, 'A, 'A tree).

declare fun sum(i32 tree) : i32.
fun sum(Tree) =
    match Tree with
    | leaf => 0
    | node(L, V, R) => V + sum(L) + sum(R)
    end.

declare input num_tree(i32 tree).
num_tree(node(leaf, 42, leaf)).
num_tree(node(node(leaf, 1, leaf), 3, node(leaf, 5, leaf))).

declare output tree_sum(i32 tree, i32).
tree_sum(Tree, Sum) :- num_tree(Tree), sum(Tree) = Sum.
    \end{verbatim}
    \caption{A \Name program consists of type-related metadata and the definitions of functions and relations.}
    \label{fig:tree}
\end{figure}

We use a toy program to help introduce \Name's use of types, constructors, functions, and relations (Figure~\ref{fig:tree}).
A \Name program conceptually consists of two parts: type-related metadata, such as type definitions and function type signatures, and program logic, such as function and rule definitions.
The sample program begins with the definition of an abstract data type \icode{tree}, which is parameterized by a type variable \icode{'A}.
In addition to algebraic data types, \Name supports parameterized tuple types, and has  
 base types \icode{string} and \icode{i32} (32-bit signed integers).
Tuples and the constructors of algebraic data types can be used as terms within \Name rules, which means that \Name has uninterpreted functions, unlike Datalog, which only has uninterpreted constants. 

Functions, such as \icode{sum} in the example, are declared with a (possibly polymorphic) type signature and defined using an ML-style syntax that includes two kind of expressions: plain terms and match expressions (with syntactic sugar for \icode{let} and \icode{if} expressions).
As functions consume and produce ground terms, every variable in a function must be bound by an outer context.
Functions in \Name are not higher-order.
Functions can be (mutually) recursive, and it is the programmer's responsibility to ensure termination.

Relations are also declared with a type signature.
The keywords \icode{input} and \icode{output} label, respectively, EDB relations (like \icode{num\_tree}) and IDB relations (like \icode{tree\_sum}).
Relations are defined as in Datalog.
In the example, the \icode{num\_tree} EDB relation is tabulated explicitly by two facts.
The IDB relation \icode{tree\_sum} is defined by a single rule
(a premise of the form \icode{$s$ = $t$} is true whenever $s$ and $t$ can be unified).
The invocation of the function \icode{sum} appears as a normal term in the rule.
The sample program generates two facts for the \icode{tree\_sum} relation:
\begin{verbatim}
    tree_sum(node(leaf, 42, leaf), 42)
    tree_sum(node(node(leaf, 1, leaf), 3, node(leaf, 5, leaf)), 9)
\end{verbatim}

\subsection{Supporting logical formulae}\label{sec:formulae}

A \Name program represents logical formulae using ground terms, manipulates formulae with user-defined functions, and reasons about formulae through built-in SMT support.
We discuss each in turn.

\subsubsection{Representing logical formulae}

\begin{figure}[!t]
    \begin{minipage}{.5\textwidth}
        \begin{verbatim}
define type bool_exp =
    | true
    | false
    | not(bool_exp)
    | and(bool_exp, bool_exp)
    | or(bool_exp, bool_exp)
    | bv32_eq(bv32_exp, bv32_exp)
    | bv32_slt(bv32_exp, bv32_exp)
    | bv32_sgt(bv32_exp, bv32_exp).
        \end{verbatim}
    \end{minipage}%
    \begin{minipage}{.5\textwidth}
        \begin{verbatim}
define type bv32_exp =
    | bv32_const(i32)
    | bv32_sym(string)
    | bv32_neg(bv32_exp)
    | bv32_add(bv32_exp, bv32_exp)
    | bv32_sub(bv32_exp, bv32_exp)
    | bv32_div(bv32_exp, bv32_exp)
    | bv32_mul(bv32_exp, bv32_exp)
    | bv32_rem(bv32_exp, bv32_exp).
        \end{verbatim}
    \end{minipage}
    \caption{Logical formulae are represented in \Name using constructors defined in the algebraic data types \icode{bool\_exp} and \icode{bv32\_exp}.}
\label{fig:formulae}
\end{figure}

\Name currently has support for constructing formulae concerning booleans and 32-bit bit vectors,
although it would be entirely possible to extend the formula language to support reasoning about floating point numbers, arrays, etc. 
These formulae are created using the constructors belonging to two built-in algebraic data types (Figure~\ref{fig:formulae}). 
The type \icode{bool\_exp} represents normal logic connectives and (signed) comparisons between 32-bit bit vector expressions, which are represented by the type \icode{bv32\_exp}.
The constructor \icode{bv32\_const(N)} takes a 32-bit integer as an argument and represents a bit vector constant of value \icode{N}.
The constructor \icode{bv32\_sym(S)} represents a symbolic bit vector identified by a string \icode{S}.
For example, the term
\begin{verbatim}
    and(bv32_eq(bv32_sym("x"), bv32_const(42)),
        bv32_eq(bv32_sym("y"), bv32_add(bv32_sym("x"), bv32_const(1))))
\end{verbatim}
represents the constraint \icode{x = 42 $\wedge$ y = x + 1}, where \icode{x} and \icode{y} are bit vectors.

\subsubsection{Manipulating logical formulae}

Because formulae are normal terms, they can be manipulated using standard functions.
For instance, this function performs a symbol substitution in a bit vector expression:
\begin{verbatim}
    declare fun subst(string, string, bv32_exp) : bv32_exp.
    fun subst(Old, New, E) =
        match E with
        | bv32_const(N) => E
        | bv32_sym(S)   => if S == Old then bv32_sym(New) else E
        | bv32_neg(E1)  => bv32_neg(subst(Old, New, E1))
        ...
        end.
\end{verbatim}
The infix function \icode{==} returns true if its two arguments reduce to the same term.

\subsubsection{Reasoning about logical formulae}

\Name provides built-in functions that support reasoning about logical formulae.
The function \icode{is\_sat($t$)} takes as an argument a term of type \icode{bool\_exp} and returns a boolean indicating whether that term is satisfiable.
For example, say that the variable \icode{Z} is bound to the formula presented above that represents the constraint \icode{x = 42 $\wedge$ y = x + 1}.
In this case, \icode{is\_sat(Z)} evaluates to true, since there exist values for \icode{x} and \icode{y} that make the formula true when interpreted under the theory of bit vectors (namely, 42 and 43).
On the other hand, the invocation \icode{is\_sat(and(Z, false))} evaluates to the boolean false.
Our current prototype also includes a function for computing Craig interpolants~\cite{craig}, which are useful for abstract model checking~\cite{lazyabsinterpolants}, and we anticipate adding support for additional logical queries, such as finding a model for a formula.

\subsection{Semantics}\label{sec:semantics}

Although we have not yet formally established the semantics of \Name, we conjecture that the meaning of a well-formed \Name program is the least model of the theory that corresponds to the program.
This would give it a semantics similar to that of a Datalog program.
However, this model could differ from a model corresponding to a Datalog program in two significant respects.
First, since \Name has $n$-ary constructors, the least model might be infinite.
Currently it is the programmer's responsibility to ensure that the model is finite, but we hope to develop analyses that will warn the programmer if it appears that the model might be infinite (while not issuing too many false alarms).
Second, since \Name has interpreted functions, the model would depend on the meaning of the functions in the program and would not be a Herbrand model.
This need not necessarily detract from the declarativeness of \Name:
as long as every function is pure and evaluates to a ground term,
then every function has a mathematical meaning that is independent of its evaluation.
For example, the meaning of a pure function does not depend on whether a call-by-value or call-by-name reduction strategy is used.
As a well-formed \Name program is composed of declarative parts, we would expect the program as a whole to have a declarative semantics.
In particular, the meaning of a well-formed program should be independent of the order of rules and the order of atoms within rule bodies.

\begin{figure}[!t]
    \begin{verbatim}
define type node = i32.
define type reg = string.
define type cond = cond_eq | cond_ne | cond_lt | cond_le | ...
define type inst = inst_jmp(cond, reg, reg, node) | inst_fail | ...
declare input stmt(node, inst).
declare input fall_thru_succ(node, node).

define type store = (reg, bv32_exp) map.
define type state = (store * bool_exp * i32 * i32).
declare input init_fuel(i32).
declare input start(node, store).
    \end{verbatim}
    \caption{Program instructions can be represented using algebraic data types and a CFG can be represented using EDB relations.
    The symbolic executor uses complex terms to keep track of program state; initial state is set by the value of EDB relations.}
    \label{fig:symexec_repr}
\end{figure}

\section{Case study: symbolic execution}\label{sec:cases}

This section demonstrates how \Name can be used to implement symbolic execution.
The encoding is natural and relatively concise (about 200 lines of code),
and leverages all of the language features described in the previous section.

A symbolic executor interprets a program in which some values are symbolic, in that they represent not a single runtime value but a set of runtime values~\cite{king,klee,symexecsurvey}.
Typically, the set associated with a value is initially unconstrained, meaning that it can concretely have any value of the relevant type.
However, when the symbolic executor reaches a condition that depends on a symbolic value, it will fork into two processes, one in which the condition is true and one in which the condition is false.
In each of the forks, it constrains the symbolic value so that it is consistent with the branch that has been taken.
For example, say that a variable \icode{x} is associated with the unconstrained symbolic value $\alpha$.
When execution encounters the branch \icode{x < 42}, execution will split and explore both branches.
However, in the ``then'' branch $\alpha$ will be constrained to be less than 42, and in the ``else'' branch it will be constrained to be greater than or equal to 42.
By checking to make sure that the accumulated constraints on a symbolic value are consistent, execution can avoid exploring infeasible paths through the program, which saves analysis time and reduces false positives.

Our \Name implementation of symbolic execution takes as input a control flow graph (CFG) representation of a simple register language (Figure~\ref{fig:symexec_repr}, top half).
The input relation \icode{stmt} relates a CFG node to an instruction in this language.
Although our full implementation supports a wider range of instructions (such as unary and binary arithmetic operations), here we focus on two instructions: a conditional jump instruction that steps to a node based on the result of comparing two registers, and a fail instruction, which represents that execution has reached a failed assertion and a bug has been found.
The relation \icode{fall\_thru\_succ} relates a node to its non-jump successor in the CFG.

\begin{figure}[!t]
    \begin{verbatim}
declare output reach(node, state).
declare output step_to(node, state).

reach(Node, New_state) :-
    step_to(Node, State),
    decr_fuel(State) = some(New_state).

step_to(Succ, New_state) :-
    reach(Node, State),
    stmt(Node, inst_jmp(Cond, Val1, Val2, Succ)),
    New_state = add_cond_to_state(Cond, Val1, Val2, State),
    is_sat(get_constraints(New_state)) = true.

declare fun add_cond_to_state(cond, reg, reg, state) : state.
fun add_cond_to_state(Cond, Reg1, Reg2, State) =
    let (Store, Constraints, Count, Fuel) = State in
    let some(Val1) = get(Reg1, Store) in
    let some(Val2) = get(Reg2, Store) in
    let Constraint =
        match Cond with
        | cond_eq => bv32_eq(Val1, Val2)
        | cond_ne => not(bv32_eq(Val1, Val2))
        ...
        end in
    (Store, and(Constraint, Constraints), Count, Fuel).

declare output failed_assert(node, state).
failed_assert(Node, State) :-
    reach(Node, State),
    stmt(Node, inst_fail).

    \end{verbatim}
    \caption{The evaluation of the symbolic executor is defined using a combination of relations and functions.}
    \label{fig:symexec_interp}
\end{figure}

The symbolic interpreter needs to maintain state that tracks  
the current value of each register and any constraints on symbolic values (Figure~\ref{fig:symexec_repr}, bottom half).
Accordingly, the type \icode{state} consists of a \icode{store} (a map from registers to \icode{bool\_exp}, implemented as an association list) and a \icode{bool\_exp} that tracks constraints on the program path.
The state also includes two integers.
The first exists for a purely technical reason (to create fresh symbols).
The second records the amount of ``fuel'' left for execution.
Each step in the symbolic execution consumes a unit of fuel.
This guarantees that execution will terminate, but also means that the analysis is in general unsound and may miss some bugs.
The amount of fuel to use is set by the EDB relation \icode{init\_fuel}, and the EDB relation \icode{start} sets the entry point of the CFG and the initial store.

Two mutually recursive relations define the symbolic interpreter (Figure~\ref{fig:symexec_interp}).
The relation \icode{reach} records that execution has reached a CFG node with a particular execution state, while
the relation \icode{step\_to} records that execution is attempting to take a step to a particular node.
The relation \icode{reach} is defined by a single rule, which says that a node is reachable if execution is trying to step to it and the execution state has enough fuel (the function \icode{decr\_fuel} returns the option constructor \icode{none} if the state is out of fuel).
On the other hand, there is a different \icode{step\_to} rule for each possible execution step.
The rule we give here describes performing a conditional jump.
The function \icode{add\_cond\_to\_state} adds a new constraint on the execution path representing the condition for the jump, and the function \icode{is\_sat} is used to make sure that the condition is actually satisfiable given the state's accumulated constraints.
Another rule would be necessary to handle the case when the negation of the condition is satisfiable.
Since these cases are not mutually exclusive, both rules can be triggered simultaneously, which corresponds to the symbolic executor forking into multiple processes.
Finally, the rule \icode{failed\_assert} is triggered when a fail instruction is reachable.

It was very straightforward to implement the symbolic interpreter presented in this section.
Of course, it is for a toy language, and
in future work we hope to evaluate whether \Name allows a similarly clean encoding of symbolic execution for a more complex language, like JVM bytecode.
We also hope to test whether an analysis written in \Name can be sufficiently performant.
For example, successful symbolic execution engines typically use heuristics to focus on high-priority areas of the execution state space.
It would be interesting to see if these heuristics could be encoded in \Name and, if not, what sort of language features could be added to \Name to make this possible while still keeping the language declarative.

In addition to the symbolic interpreter described in this section, we have implemented an abstract model checker that operates over a similar register language.
The model checker is based on the lazy abstraction paradigm~\cite{lazyabs} and uses interpolants to refine abstractions~\cite{lazyabsinterpolants}.
The implementation is about 500 lines of \Name.

\section{Prototype}\label{sec:prototype}

We have written a prototype implementation of \Name in approximately 6,000 lines of Java code.
Here we outline the system, give some initial performance numbers, and propose some potential optimizations.

\subsection{System design}

The \Name prototype works in five stages: parsing, type checking, program validation, program rewriting, and evaluation.
\Name source code is first parsed into an abstract syntax tree (AST) representation using a generated parser.
The AST is then type checked using a unification-based algorithm.
This involves type checking the rules, facts, and function definitions against the type signatures declared in the program metadata.
All type information is discarded after type checking.
Next, a validator checks that the AST represents a well-formed program.
In particular, it ensures that the program meets the requirements of stratified negation and some restrictions on the use of variables.
In addition to the standard range restriction of Datalog, \Name requires that every variable that appears within an argument to a function is bound elsewhere in the clause. This is necessary since functions in \Name are not invertible in general.
Finally, a rule preprocessor reorders the premises in each rule so that the premises can be evaluated from left to right, ensuring that a variable is bound before being used in a function call.
Rewriting rules in this way is only safe if every function is pure (which, in the context of \Name, means that every function must evaluate to a ground term).

The \Name interpreter evaluates one stratum of the program at a time.
Each stratum is evaluated using a parallelized bottom-up algorithm inspired by pipelined semi-naive evaluation~\cite{network}.
Parallelism is effected through a work-stealing thread pool.
Each work item is a rule that has been partially evaluated on a fact; a worker thread completes a work item by fully evaluating the rule against a database of currently derived facts.
When evaluation results in a fact that has not been seen before, each relevant rule is partially evaluated on that fact and then submitted as a new work item.
When a logical function like \icode{is\_sat} is invoked in the course of evaluation, the arguments are translated into Z3 formulae and a query is made to the Z3 SMT solver.%
\footnote{Z3 is available at \url{https://github.com/Z3Prover/z3}.}

\subsection{Initial performance results}

We have tested the performance and scalability of our prototype on two programs and,
while our prototype was not exceptionally fast, it did achieve a respectable level of parallel scaling in both cases.
All measurements were taken on 2.5 GHz Intel Xeon Platinum 8175 machine running Ubuntu 16.04, with 192 GiB of memory and 24 hyper-threaded physical cores.
We varied the size of the work-stealing thread pool from one to 48 threads by multiples of eight.
We performed three trials for each test configuration, and report the median result here.

\begin{table}
    \caption{While not exceptionally fast, the \Name prototype achieved respectable parallel scaling on two test cases. This figure shows absolute times, speedup relative to using a single thread, and efficiency (the ratio of speedup to the number of threads).}
    \label{tab:results}
    \begin{minipage}{\textwidth}
        \begin{tabular}{llccccccc}
            \hline\hline
            & & \multicolumn{7}{c}{Number of threads} \\
            \noalign{\vspace {.1cm}}
            & & 1 & 8 & 16 & 24 & 32 & 40 & 48 \\
            \hline
            & Time (s) & 303 & 49 & 27 & 24 & 21 & 19 & 18 \\
            Transitive closure\quad\quad & Speedup  & 1.0 & 6.2 & 11.2 & 12.6 & 14.4 & 15.9 & 16.8 \\
            & Efficiency  & 1.00 & 0.77 & 0.70 & 0.53 & 0.45 & 0.40 & 0.35 \\
            \hline
            & Time (s) & 90 & 17 & 12 & 10 & 10 & 9 & 9 \\
            Symbolic execution\quad\quad & Speedup  & 1.0 & 5.3 & 7.5 & 9.0 & 9.0 & 10.0 & 10.0 \\
            & Efficiency  & 1.00 & 0.66 & 0.47 & 0.38 & 0.28 & 0.25 & 0.21
        \end{tabular}
    \end{minipage}
\end{table}

The first test program computed the transitive closure of a random directed graph (Table~\ref{tab:results}, top).
The program was written in a pure-Datalog fragment of \Name and never invoked the SMT solver.
The graph had 2,000 vertices, and each potential edge was included in the graph with a probability of 0.1.
This very dense graph was strongly connected, so that the transitive closure relation had four million tuples.
Performance improved with the size of the thread pool, although the speedup was sublinear.
Using 48 threads, \Name computed the transitive closure relation in 18 seconds, a speedup of 16.7x relative to using a single thread.
For comparison, Souffl\'e~\cite{souffle2} computed the transitive closure relation in about two seconds while achieving a speedup of around 3x (using 48 threads versus a single thread); on the other hand, even with 48 threads Flix~\cite{flix} did not finish in a reasonable amount of time.

The second test ran the symbolic execution implementation from Section~\ref{sec:cases} (Table~\ref{tab:results}, bottom).
It achieved a max speedup of 10x, when it was able to explore over 5,800 program states in about nine seconds.
The symbolic interpreter was invoked on an input program with a high branching factor (a tight loop in the program forces the symbolic interpreter to fork into three processes each iteration).
Consequently, the symbolic executor had to make a call to Z3 at nearly every step.
We conjecture that this test achieved a lower speedup relative to the transitive closure example because of the higher percentage of work done outside of our parallel algorithm, such as in interprocess communication and constraint solving.

\subsection{Future optimizations}

The performance of our prototype has been acceptable on small toy examples, but we anticipate that making it scale will require some smart optimizations.
We have identified five main areas that we would like to focus on: memoization, SMT solving, database optimizations, work stealing, and unification.

\subsubsection{Memoization}
Our prototype memoizes terms and atoms during creation, so that there are never multiple Java objects that represent terms or atoms that are syntactically the same.
        The advantage to this approach is that the equality of two terms (or atoms) and the hash code of a term (or atom) can be computed in constant time.
        Since these operations happen often during rule evaluation, we expect memoization to have a net benefit on performance, outweighing the downside that a shared cache needs to be checked every time a term or atom is created.
        However, more experimentation is necessary to determine the impact of memoization on real workloads.
        It would also be possible to extend memoization to other parts of the runtime, such as caching the results of function calls (which is possible because functions are pure and deterministic).

\subsubsection{SMT solving}
Making calls to an SMT solver can be expensive and applications that depend on SMT solving benefit by limiting the number of calls they make.
        For instance, the symbolic execution engine KLEE caches SMT query results such that, in certain cases, it is possible to tell if the result of one SMT query implies another without having to make another SMT call~\cite{klee}.
        Our prototype currently uses Z3 naively.

\subsubsection{Database optimizations}
Datalog implementations that can handle industry-scale static analyses depend on database-style optimizations, such as optimizing the order that premises in a rule body are evaluated (akin to optimizing the order of database operations) and using indexed data structures to store Datalog facts~\cite{doop,souffle1,souffle2}.
        Our \Name implementation does not optimize the order in which it evaluates rule premises and,
        while it does use an indexed data structure to store facts, this data structure has not been extensively optimized.

\subsubsection{Work stealing}
We have not experimented with the size of the work items we submit to the work-stealing thread pool. However, we anticipate having to tune the granularity of work items to get optimal performance out of our parallel evaluation algorithm.

\subsubsection{Unification}
We have developed an algorithm to unify two terms that runs in time linear in the number of subterms (ignoring the time necessary to reduce any functions occurring in the terms).
        This is tricky in the presence of functions, which can only be evaluated after every variable in their arguments is bound.
        For example, the calls to the function \icode{f} in this example force the subterms to be unified in a particular order (\icode{a} and \icode{b} are constructors):
\begin{align*}
    &\icode{a\big(f(X$_n$), f(X$_{n-1}$), ..., f(X$_0$), b(0)\big) =} \\
    &\icode{a\big(b(X$_{n+1}$), b(X$_n$), ..., b(X$_1$), b(X$_0$)\big)}
\end{align*}
        \icode{b(X$_0$)} must first be unified with \icode{b(0)}, then \icode{b(X$_1$)} can be unified with the result of \icode{f(X$_0$)}, then \icode{b(X$_2$)} can be unified with the result of \icode{f(X$_1$)}, and so on.
        A naive algorithm might make a quadratic number of left-to-right passes over the terms.
        Although our linear-time algorithm should have performance benefits in theory, in practice a naive algorithm might outperform it on typical unification instances because of constant factors.
        As an alternative approach, complex unification instances (such as the one above) could be rewritten into a collection of simpler unification instances, which in turn could be solved with a simpler algorithm.
        As unification is such a common operation in \Name evaluation, the performance of the unification algorithm could have a substantial impact on overall performance.

\section{Conclusion}\label{sec:conclusion}

\Name is an extension of Datalog that makes it possible to represent, manipulate, and reason about logical formulae.
In turn, this enables one to declaratively implement static analyses such as symbolic execution and abstract model checking.
While we anticipate that major engineering work will be required to scale \Name to real-world static analysis problems,
we nonetheless believe that languages like it have the potential to help address some concrete challenges faced by modern static analyses.

\bibliographystyle{acmtrans}
\bibliography{paper}

\label{lastpage}
\end{document}